\documentclass[]{aa}

\usepackage{graphics}

\begin{document}

   \thesaurus{08         
              (08.01.1;  
               08.01.3;  
               08.08.2;  
               10.15.2)} 
   \title{Chemical composition of evolved stars \\ 
in the open cluster M~67\thanks{Based on observations obtained at 
the Nordic Optical Te\-les\-co\-pe, La Palma}}

   \author{G. Tautvai\v{s}ien\.{e}\inst{1},
   B. Edvardsson\inst{2},
   I. Tuominen\inst{3} and I. Ilyin \inst{3}} 

   \institute{$^1$ Institute of Theoretical Physics and Astronomy,
              Go\v{s}tauto 12, Vilnius 2600, Lithuania\\
              $^2$ Uppsala Astronomical Observatory, Box 515, 
	      SE-751\thinspace20 Uppsala, Sweden\\ 
              $^3$ Astronomy Division, Dept. of Physical Sciences, 
              University of Oulu, P.O. Box 300, 90401, Oulu, 
              Finland
             }

   \authorrunning{G. Tautvai\v{s}ien\.{e} et al.}
   \titlerunning{Chemical composition of evolved stars in M~67}

   \date{Received January 00, 2000; accepted January 00, 2000}
   \maketitle
   
   \begin{abstract}
High-resolution spectra of six core helium-burning `clump' stars and three 
giants in the open cluster M~67 have been obtained with the SOFIN spectrograph 
on the Nordic Optical Telescope to investigate abundances of up to 25 
chemical elements. Abundances of carbon were studied using the ${\rm C}_2$ 
Swan (0,1) band head at 5635.5~{\AA}.  The wavelength interval 
7980--8130~{\AA} with strong CN features was analysed in order to determine 
nitrogen abundances and $^{12}{\rm C}/^{13}{\rm C}$  isotope ratios. 
The oxygen abundances were determined from the [O~I] line at 6300~{\AA}. 

The overall metallicity of the cluster stars was found to be close 
to solar ([Fe/H]=$-0.03\pm0.03$).  Compared with the Sun and other dwarf 
stars of the Galactic disk, as well as with dwarf stars of M~67 itself, 
abundances in the investigated stars suggest that carbon is depleted 
by about 0.2~dex, nitrogen is enhanced by about 0.2~dex and oxygen is 
unaltered. Among other mixing-sensitive chemical elements an overabundance 
of sodium may be suspected. The mean C/N and $^{12}{\rm C}/^{13}{\rm C}$ 
ratios are lowered to the values of $1.7\pm0.2$ and $24\pm4$ in the giants 
and to the values of $1.4\pm0.2$ and $16\pm4$ in the clump stars. 
These results suggest that extra mixing of CN-cycled material to the
stellar surface takes place after the He-core flash.
Abundances of heavy chemical elements in all nine stars were 
found to be almost identical and close to solar. 

   \end{abstract}

\section{Introduction}

The old open cluster M~67 has served as an important sample in 
understanding stellar evolution of almost fifty years. Since the first papers 
by Becker \& Stock (1952), Popper (1954), Johnson \& Sandage (1955) work has 
continued and more than 200 studies were accomplished (see, e.g., 
Burstein et al.\ 1986, Carraro et al.\ 1996 and references therein). The 
advantage that cluster members have to be coeval and identical except for 
mass and evolutionary state, which can be identified unambiguously, may most 
efficiently serve for the analysis of changes in mixing-sensitive abundances. 

Abundances of carbon and nitrogen are particularly sensitive tests for  
stellar evolution. The enhancement of CN bands was reported for the 
M~67 clump stars F84, F141 and F151 already by Pagel (1974). 
From high-resolution spectra, C and N abundances have been investigated 
in three giants (Brown 1985) and from moderate-resolution spectra in 19 
giants (Brown 1987). Carbon isotope ratios along the giant branch 
were investigated by Gilroy (1989) and Gilroy \& Brown (1991). It was found 
that C/N and $^{12}{\rm C}/^{13}{\rm C}$ ratios in the clump giants and the 
stars at the tip of the giant branch all have values much lower than predicted 
in standard models. Charbonnel et al.\ (1998), however, suggest that the 
absolute values of C/N ratios obtained by Brown (1987) may have a systematic 
offset and have to be taken with caution. 

High-resolution analyses are very scarce for the oxygen
abundances: Griffin (1975) has measured the [O I] line at 6300~{\AA} in one
star, but this star (IV-202) is quite cool and the result is uncertain; Cohen
(1980) has analysed four stars, but the weaker line of [O I] at 6363~{\AA} 
was used. In the paper by Cohen (1980), a low value of [Fe/H]$=-0.39$ for M~67 
has been received. This is the same paper that gave a low [Fe/H] value for 
the globular cluster M~71, which was later increased by +0.5~dex in Cohen 
(1983). The same correction, if applied to M~67, would yield the [Fe/H] near 
solar, consistent with more recent determinations (e.g.\ Nissen et al.\ 1987,
Garcia Lopez et al.\ 1988, Hobbs \& Thorburn 1991, Friel \& Boesgaard 1992). 
In the same paper by Cohen (1980), which is based on photographic data as well 
as all M\thinspace67 abundance works before Brown (1985), abundances of some 
other elements in M~67 look very extraordinary. The ratios of [Mg/Fe] 
reach --0.8~dex while the ratios of other $\alpha$-process element [Si/Fe] 
are enhanced by about +0.6~dex, [Ba/Fe] are approximately equal to --0.4~dex 
while for the very similar element lanthanum [La/Fe]$\approx$+0.6~dex.      
    
In this paper, we report a detailed analysis of six core 
helium-burning clump stars and three giants in M~67 (see Fig. 1 for their 
location in the HR diagram). The core He-burning stars are the
most evolved stars in M~67, their  surface abundances reflect effects of the 
preceding evolution along the red giant branch as well as effects raised by 
the helium flash. The study aims at a very high internal precision of the 
abundances, so that even small anomalies in the chemical composition can be 
revealed.   

  \begin{figure}
    \resizebox{\hsize}{!}{\includegraphics{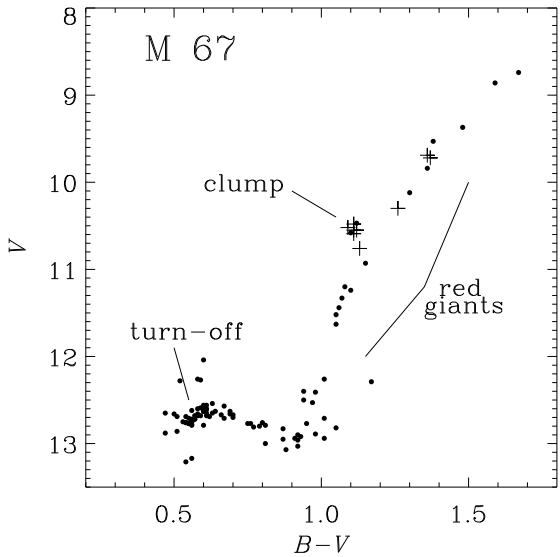}}
    \caption
    {The colour-magnitude diagram of the open cluster M~67. The red 
    giants and the core He-burning `clump' stars analysed are indicated by the 
    crosses. The diagram is taken from Mathieu et al. (1986) and cleaned 
    from evident binary stars}
    \label{CMD}
  \end{figure}

\section{Observations and data reductions}
  The spectra were obtained at the Nordic Optical Telescope (NOT) with 
the SOFIN \'{e}chelle spectrograph (Tuominen 1992) in February of 1995.  
The 3rd optical camera was used with an entrance slit width of $197~\mu$m,
providing a spectral resolving power of $R1\approx 30\thinspace000$.
Three stars 
(F84, F141 and F151) were also observed with the 2nd optical camera, providing 
a spectral resolving power of $R2\approx 60\thinspace000$ (the entrance slit
width $68~\mu$m). The spectra were recorded with the 
Astromed-3200 CCD camera (Mackay 1986) equipped with an EEV P88100 UV-coated 
CCD of $1152\times298$ pixels operated at the working temperature of 150 K. 
With the 3rd camera the CCD size covered simultaneously 25 spectral orders, 
each of 80--150~{\AA} in length, located in the interval of 4500-8750~{\AA}. 
The 2nd camera gave 15 spectral orders, each of 40--60~{\AA}, in the spectral 
region  from 5650 to 8750~{\AA}. For the F84 and F141 
two partly overlapping spectral regions were 
observed by setting of the \'{e}chelle and cross-dispersion prism angles in 
order to extend the wavelength intervals of orders. The concrete numbers of 
spectrograms received for every star are presented in Table 1.   
All spectra were exposed to S/N$\ge 100$.  

Reductions of the CCD images were made with the {\it 3A} software package 
(Ilyin 1996). Procedures of bias subtraction, spike elimination, flat field 
correction, scattered light subtraction, extraction of spectral orders were 
used for image processing. A Th-Ar comparison spectrum was used for the 
wavelength calibration. The continuum was defined by a number of narrow 
spectral regions, selected to be free of lines. 

The lines suitable for 
measurement were chosen using the requirement that the profiles be 
sufficiently clean to provide reliable equivalent widths. Inspection of the 
solar spectrum (Kurucz et al.\ 1984) and the solar line identifications of 
Moore et al.\ (1966) were used to avoid blends. Lines blended by 
telluric absorption lines were omitted from treatment as well. 
The equivalent widths of lines were measured by fitting of a Gaussian 
function. 179 atomic lines were selected for the analysis in spectra of the 
3rd camera and 132 in spectra obtained on the 2nd camera. The line 
measurements are listed in Tables 2 and 3 (available in electronic form at 
CDS). 
    

\section{Method of analysis and physical data}
The spectra were analysed using a differential model atmosphere technique. 
The {\it Eqwidth} and {\it Synthetic Spectrum} program packages, developed 
at the Uppsala Astronomical Observatory, were used to carry out 
the calculations of theoretical equivalent widths of lines and synthetic 
spectra. A set of plane parallel, line-blanketed, flux constant LTE model 
atmospheres was computed 
with an updated version of the {\it MARCS} code of Gustafsson et al.\ (1975) 
using continuous opacities from Asplund et al.\ (1997) and including UV line 
blanketing as described by Edvardsson et al.\ (1993). Convection was treated 
in the mixing-length approximation ($l/H_p=1.5$). 

The Vienna Atomic Line Data Base (VALD, Piskunov et al.\ 1995) was extensively 
used while preparing the input data for the calculations. Atomic oscillator 
strengths for this study were taken mainly from two sources: the first 
being an inverse solar spectrum analysis done in Kiev (Gurtovenko \& Kostik 
1989, Gurtovenko et al.\ 1983, 1985a, 1986), the second being high-precision 
laboratory measurements done in Oxford (Blackwell et al.\ 1982, 1983, 1986). 
The coincidence of these two sets of $gf$ values is very good, and the errors 
in the least-squares fit do not exceed $\pm~0.07$~dex (Gurtovenko et al.\ 
1985b). For Ca~I the $gf$ values have been also taken from  
Smith \& Raggett (1981) and for Zr~I from Bogdanovich et al.\ (1996). 

Using the $gf$ values and solar equivalent widths of analysed lines from the 
cited sources we have obtained the solar abundances, later used for the 
differential determination of abundances in the programme stars. We used the 
solar model atmosphere from the set calculated in Uppsala (Edvardsson et al.\ 
1993) with a microturbulent velocity of 0.8~$\rm {km~s}^{-1}$, as derived from 
Fe~I lines. 

Abundances of carbon, nitrogen and europium were determined using the spectrum 
synthesis technique. 
The interval of 5632--5636~{\AA} was synthesized and 
compared with observations in the vicinity of the ${\rm C}_2$ Swan 0--1 band 
head at 5635.5~{\AA}. The same atomic data of ${\rm C}_2$ as used by 
Gonzalez et al. (1998) were adopted for the analysis. 
The interval of 7980--8130~{\AA}, containing strong CN features, 
was analysed in order to determine the nitrogen abundance and 
$^{12}{\rm C}/^{13}{\rm C}$ ratios. 
The molecular data for
$^{12}{\rm C}^{14}{\rm N}$ and $^{13}{\rm C}^{14}{\rm N}$
were taken from {\it ab initio} calculations of CN isotopic line strengths,
energy levels and wavelengths by Plez (1999), with all $gf$ values increased
by +0.03 dex in order to fit our model spectrum to the solar atlas of
Kurucz et al.\ (1984).
Parameters of other lines in the intervals 
of spectral synthesis were compiled from the VALD database. 
In order to check the correctness of the 
input data, synthetic spectra of the Sun were compared to the 
solar atlas of Kurucz et al.\ (1984) and necessary adjustments were made
to the line data.      

An interval of 6643--6648~{\AA}, containing the Eu~II line at 6645~{\AA}, was 
analysed in order to determine the europium abundance. 
The oscillator strength of the Eu~II line , log~$gf$=0.17, 
was adopted from Gurtovenko \& Kostik (1989). The solar abundance of 
europium, later used for the differential analysis, 
log$A({\rm Eu})_{\odot}$=0.49, was determined using the same procedure as 
for other heavy chemical elements. Parameters of other lines in the interval 
were compiled from the VALD database. CN lines were also included, but none 
of them seems to affect the europium line significantly. 
 
In addition to thermal and microturbulent Doppler broadening of lines, atomic 
line broadening by radiation damping and van der Waals damping were considered 
in the calculation of abundances. Radiation damping parameters for the most of 
lines were taken from the VALD database. When they were not available at the 
VALD database, published oscillator strengths of strong lines were used for 
determination of life times and thus radiation damping for relevant energy 
levels. Correction factors to the classical van der Waals damping widths 
$(\Gamma_6)$ were taken from the literature: Na~I: Holweger (1971), Ca~I: 
O'Neill \& Smith (1980), Ba~II: Holweger \& M${\rm \ddot{u}}$ller (1974), 
Fe~I: Simmons \& Blackwell (1982). For all other species a correction factor 
of 2.5 was applied to the classical $\Gamma_6$ $(\Delta {\rm log}C_6$=+1.0), 
following M${\rm \ddot{a}}$ckle et al.\ (1975). For lines stronger than 
$W$=100~m{\AA}, the correction factors were selected individually by the 
inspection of the solar spectrum. 

\section{Atmospheric parameters and abundances}

The  effective  temperatures were derived and averaged from the intrinsic 
colour indices ({\it B--V})$_0$ and ({\it V--K})$_0$ using the corresponding 
calibrations by Gratton et al.\ (1996). Colour indices have been taken from 
Houdashelt et al.\ (1992) and dereddened using $E_{B-V}$=0.032 according to 
Nissen et al.\ (1987) and $E_{B-V}/E_{V-K}$=3.24 according to Taylor \& Joner 
(1988). For F266 the {\it B--V} index was taken from 
Coleman (1982) and {\it V--K} from Taylor \& Joner (1988).  
The agreement between the temperatures deduced from the two 
colour indices is quite good, the differences do not exceed 20~K. The 
gravities were found by forcing  Fe~I and Fe~II to yield the same iron 
abundances. The microturbulent velocities were determined  by
forcing Fe~I line abundances to be independent of the equivalent width. 
The derived atmospheric parameters are listed in Table~1. 

   \begin{table}
      \caption{Atmospheric parameters derived for the programme stars. 
The last two columns give the resolving powers and number of spectra observed }
      \[
         \begin{tabular}{lccrccc}
            \hline
            \noalign{\smallskip}
 Star & $T_{\rm eff}~{\rm (K)}$ & 
            log~$g$ & [Fe/H] & 
            ${ v_{\rm t}~{\rm (km~s}^{-1}}$)&R$^*$&N \\ 
            \noalign{\smallskip}
            \hline
            \noalign{\smallskip}
 F84     & 4750 & 2.4 & --0.02 & 1.8 &$R1$ & 3 \\  
         & 4750 & 2.4 & --0.05 & 1.6 &$R2$ & 3 \\  
 F105    & 4450 & 2.2 & --0.05 & 1.9 &$R1$ & 2 \\  
 F108    & 4250 & 1.7 & --0.02 & 1.8 &$R1$ & 3 \\  
 F141    & 4730 & 2.4 & --0.01 & 1.8 &$R1$ & 2 \\  
         & 4730 & 2.4 &   0.01 & 1.6 &$R2$ & 4 \\  
 F151    & 4760 & 2.4 &   0.01 & 1.7 &$R1$ & 2 \\  
         & 4760 & 2.4 & --0.03 & 1.6 &$R2$ & 1 \\  
 F164    & 4700 & 2.5 &   0.00 & 1.8 &$R1$ & 2 \\  
 F170    & 4280 & 1.7 & --0.02 & 1.8 &$R1$ & 3 \\  
 F224    & 4710 & 2.4 & --0.11 & 1.6 &$R1$ & 3 \\  
 F266    & 4730 & 2.4 & --0.02 & 1.7 &$R1$ & 3 \\  
             \noalign{\smallskip}
            \hline
         \end{tabular}
      \]
\begin{list}{}{}{}{}
\item[$^*$] Resolving powers $R1\approx30\thinspace000$ and 
$R2\approx60\thinspace000$. See Sect.~3 for more explanations
\end{list}
   \end{table}

The abundances relative to hydrogen
[A/H]\footnote{In this paper we use the customary spectroscopic notation
[X/Y]$\equiv \log_{10}(N_{\rm X}/N_{\rm Y})_{\rm star} -
\log_{10}(N_{\rm X}/N_{\rm Y})_\odot$} and $\sigma$ (the line-to-line 
scatter) derived for up to 28 neutral and ionized species for the programme 
stars are listed in Tables~4 and 5. 

   \begin{table*}
  \setcounter{table}{3}
      \caption{Abundances relative to hydrogen [A/H] derived 
from spectra of $R\approx30\thinspace000$. The quoted 
errors, $\sigma$, are the standard deviations in the mean value due to the 
line-to-line scatter within the species. 
The number of lines used is indicated by $n$. }
         \label{Abund}
      \[
         \begin{tabular}{lrcrcrcrcrcrcrcrcrcr}
            \hline
            \noalign{\smallskip}
  & \multicolumn{3}{c} {F84 (clump)} &
  & \multicolumn{3}{c}{F105 (giant)} &
  & \multicolumn{3}{c}{F108 (giant)} &
  & \multicolumn{3}{c}{F141 (clump)} &
  & \multicolumn{3}{c}{F151 (clump)}\\
            \noalign{\smallskip}
\cline{2-4}\cline{6-8}\cline{10-12}\cline{14-16}\cline{18-20}
            \noalign{\smallskip}
Ion &[A/H] &$\sigma$ &$n$&\ &[A/H] &$\sigma$ &$n$&\ &[A/H]&$\sigma$&$n$ 
&\ &[A/H] &$\sigma$ &$n$&\ &[A/H]&$\sigma$&$n$\\ 
            \noalign{\smallskip}
            \hline
            \noalign{\smallskip}
C I  & --0.22&    &1 & &--0.13&    &1 & &--0.23&    &1 & &--0.18&    &1 & &--0.19&    &1 \\
N I  &   0.32&0.06&65& &  0.18&0.03&65& &  0.20&0.03&65& &  0.30&0.03&65& &  0.29&0.03&65\\
O I  &   0.03&    &1 & &  0.09&    &1 & &  0.00&    &1 & &  0.03&    &1 & &--0.05&    &1 \\
Na I &   0.17&    &1 & &  0.00&    &1 & &  0.15&    &1 & &  0.24&    &1 & &  0.22&    &1 \\
Mg I &   0.06&    &1 & &  0.02&    &1 & &  0.03&    &1 & &  0.10&    &1 & &      &    &  \\
Al I &   0.08&0.01&2 & &--0.01&0.07&3 & &  0.16&0.05&2 & &  0.07&0.01&2 & &  0.15&0.06&2 \\
Si I &   0.13&0.08&10& &  0.09&0.08&9 & &  0.03&0.10&9 & &  0.10&0.10&7 & &  0.06&0.06&7 \\
Ca I &   0.04&0.16&7 & &--0.08&0.19&7 & &  0.09&0.18&8 & &  0.08&0.12&7 & &  0.07&0.17&8 \\
Sc I & --0.06&0.11&3 & &--0.12&0.18&3 & &--0.12&0.06&2 & &--0.09&0.22&3 & &  0.03&0.23&3 \\
Sc II&   0.10&0.20&9 & &  0.10&0.21&8 & &  0.05&0.20&8 & &  0.08&0.15&9 & &  0.11&0.17&9 \\
Ti I &   0.02&0.17&21& &--0.07&0.19&22& &  0.20&0.22&22& &--0.04&0.17&20& &  0.03&0.20&22\\
Ti II&   0.12&0.13&5 & &  0.08&0.24&5 & &  0.10&0.23&5 & &  0.07&0.15&4 & &  0.10&0.16&5 \\
V I  &   0.12&0.14&17& &  0.10&0.18&17& &  0.40&0.14&14& &  0.08&0.12&17& &  0.07&0.18&16\\
Cr I &   0.08&0.11&8 & &  0.07&0.22&10& &  0.22&0.17&10& &  0.11&0.17&10& &  0.04&0.20&9 \\
Mn I & --0.10&0.12&2 & &  0.02&0.24&3 & &--0.06&0.02&2 & &  0.07&0.24&3 & &  0.05&0.23&3 \\
Fe I & --0.02&0.11&29& &--0.05&0.12&31& &--0.02&0.12&30& &--0.01&0.11&31& &  0.01&0.12&32\\
Fe II& --0.02&0.04&6 & &--0.05&0.16&6 & &--0.02&0.14&6 & &--0.01&0.12&6 & &  0.01&0.08&6 \\
Co I &   0.03&0.14&8 & &  0.11&0.17&8 & &  0.15&0.16&8 & &  0.04&0.19&8 & &  0.09&0.15&8 \\
Ni I &   0.05&0.15&22& &  0.05&0.15&22& &  0.02&0.17&22& &  0.04&0.15&22& &  0.05&0.14&21\\
Cu I &   0.13&0.06&3 & &  0.12&0.03&3 & &  0.18&0.05&2 & &  0.07&0.15&3 & &  0.12&0.12&3 \\
Y II &   0.00&0.22&4 & &  0.07&0.25&4 & &  0.11&0.18&3 & &--0.13&0.15&4 & &--0.11&0.08&4 \\
Zr I & --0.18&0.11&3 & &--0.35&0.09&3 & &--0.13&0.11&3 & &--0.19&0.19&3 & &--0.18&0.21&3 \\
Ba II&   0.06&    &1 & &--0.09&    &1 & &  0.18&    &1 & &  0.06&    &1 & &  0.14&    &1 \\
La II&   0.12&0.04&2 & &  0.20&0.01&2 & &  0.17&    &1 & &--0.07&    &1 & &  0.06&0.13&2 \\
Ce II&   0.09&0.02&2 & &  0.06&0.16&3 & &  0.04&0.15&3 & &  0.10&0.09&2 & &  0.10&0.12&2 \\
Sm II&   0.05&    &1 & &--0.04&    &1 & &      &    &  & &  0.14&    &1 & &  0.04&    &1 \\
Eu II&   0.10&    &1 & &  0.20&    &1 & &  0.00&    &1 & &  0.10&    &1 & &  0.10&    &1 \\
     &       &    &  & &           &  & &           &  & &      &    &  & &           &  \\
C/N  &   1.15&    &  & &  1.95&    &  & &  1.50&    &  &  & 1.32 &   &  & &  1.32&    &  \\ 
$^{12}$C/$^{13}$C & 20 & 6 & & & 20 & & & &21 & 3 & & & 16 & 3 & & & 17 & 3 & \\   
             \noalign{\smallskip}
   \cline{1-20}
         \end{tabular}
      \]
   \end{table*}

   \begin{table*}
  \setcounter{table}{3}
      \caption{(continued)}
      \[
         \begin{tabular}{lrcrcrcrcrcrcrcr}
            \hline
            \noalign{\smallskip}
  & \multicolumn{3}{c}{F164 (clump)} &
  & \multicolumn{3}{c}{F170 (giant)} &
  & \multicolumn{3}{c}{F224 (clump)} &
  & \multicolumn{3}{c}{F266 (clump)}\\
            \noalign{\smallskip}
\cline{2-4}\cline{6-8}\cline{10-12}\cline{14-16}
            \noalign{\smallskip}
Ion & [A/H] &$\sigma$ &$n$&\ &[A/H] &$\sigma$ &$n$&\ &[A/H]&$\sigma$&$n$ 
&\ &[A/H] &$\sigma$ &$n$\\ 
            \noalign{\smallskip}
            \hline
            \noalign{\smallskip}
C I   &--0.15&    &1 & &--0.21&    &1 & &--0.30&    &1 & &--0.21&    &1 \\
N I   &  0.24&0.03&65& &  0.16&0.03&65& &  0.10&0.03&65& &  0.18&0.03&65\\
O I   &--0.04&    &1 & &--0.05&    &1 & &--0.06&    &1 & &--0.01&    &1 \\
Na I  &  0.20&    &1 & &  0.16&    &1 & &  0.10&    &1 & &  0.20&    &1 \\
Mg I  &  0.10&    &1 & &  0.05&    &1 & &  0.07&    &1 & &  0.13&    &1 \\
Al I  &  0.16&0.02&2 & &  0.13&0.01&3 & &  0.07&0.03&2 & &  0.19&0.07&2 \\
Si I  &  0.07&0.05&9 & &  0.02&0.07&9 & &  0.07&0.05&9 & &  0.07&0.08&9 \\
Ca I  &  0.07&0.18&8 & &  0.04&0.20&7 & &--0.14&0.25&8 & &--0.03&0.19&8 \\
Sc I  &--0.16&0.19&3 & &--0.19&0.05&2 & &--0.18&0.13&3 & &--0.13&0.03&2 \\
Sc II &  0.09&0.17&9 & &  0.04&0.22&8 & &  0.00&0.19&8 & &  0.06&0.15&9 \\
Ti I  &  0.03&0.17&20& &  0.17&0.21&24& &--0.16&0.20&23& &--0.07&0.18&23\\
Ti II &  0.17&0.09&3 & &  0.17&0.16&5 & &--0.12&0.17&5 & &  0.03&0.15&5 \\
V I   &  0.13&0.13&17& &  0.27&0.17&12& &--0.06&0.14&16& &--0.06&0.12&17\\
Cr I  &  0.07&0.14&8 & &  0.11&0.20&10& &--0.07&0.22&9 & &  0.01&0.22&10\\
Mn I  &  0.11&0.17&3 & &  0.05&0.28&3 & &--0.26&0.11&2 & &--0.01&0.23&3 \\
Fe I  &  0.00&0.07&32& &--0.02&0.14&32& &--0.11&0.12&31& &--0.02&0.10&31\\
Fe II &  0.00&0.08&6 & &--0.02&0.14&6 & &--0.11&0.07&6 & &--0.02&0.06&6 \\
Co I  &  0.04&0.16&8 & &  0.08&0.23&8 & &--0.04&0.17&8 & &  0.01&0.16&8 \\
Ni I  &  0.06&0.18&22& &--0.02&0.19&22& &--0.16&0.18&23& &  0.01&0.13&22\\
Cu I  &  0.17&0.07&3 & &  0.19&0.10&2 & &--0.10&0.19&3 & &  0.04&0.07&3 \\
Y II  &  0.02&0.23&4 & &  0.14&0.14&4 & &--0.25&0.10&4 & &--0.04&0.11&3 \\
Zr I  &--0.20&0.15&3 & &--0.18&0.15&3 & &--0.11&0.23&3 & &--0.29&0.20&3 \\
Ba II &  0.02&    &1 & &  0.13&    &1 & &--0.18&    &1 & &  0.03&    &1 \\
La II &  0.17&0.02&2 & &  0.18&0.01&2 & &--0.01&0.01&2 & &  0.08&0.01&2 \\
Ce II &  0.03&0.21&2 & &  0.04&0.12&2 & &  0.02&0.10&2 & &  0.06&0.10&3 \\
Sm II &  0.08&    &1 & &  0.10&    &1 & &  0.25&    &1 & &  0.00&    &1 \\
Eu II &  0.00&    &1 & &--0.05&    &1 & &--0.15&    &1 & &  0.10&    &1 \\
      &           &  & &      &    &  & &      &    &  &  &     &    &  \\
C/N  &   1.62&    &  & &  1.70&    &  & &  1.58&    &  &  & 1.62  & & \\ 
$^{12}$C/$^{13}$C & 18 & 3 & & & 30 &7& & & 8  & 3 & & & 15 & 4 &  \\   
        \noalign{\smallskip}
   \cline{1-16}
         \end{tabular}
      \]
   \end{table*}

   \begin{table}
   \setcounter{table}{4}
      \caption{Abundances relative to hydrogen [A/H] and $\sigma$ derived 
from spectra of $R\approx60\thinspace000$. The quoted 
errors are the standard deviations in the mean value due to the 
line-to-line scatter within the species. 
The number of lines used is indicated by $n$  }
      \[
         \begin{tabular}{lrrlrlrl}
            \hline
            \noalign{\smallskip}
  & & \multicolumn{2}{c}{F84} & \multicolumn{2}{c}{F141}  &
 \multicolumn{2}{c}{F151} \\
            \noalign{\smallskip}
\cline{3-4}\cline{5-6}\cline{7-8}
            \noalign{\smallskip}
Ion & $n$ & [A/H] & $\sigma$  & [A/H] & $\sigma$  & [A/H] & $\sigma$ \\ 
            \noalign{\smallskip}
            \hline
            \noalign{\smallskip}
  O I  & 1  &   0.01 &       &  0.02 &       &  0.04  &      \\
  Na I & 3  &   0.16 & 0.09  &  0.20 & 0.15  &  0.18  &0.11  \\
  Mg I & 4  &   0.02 & 0.15  &  0.06 & 0.13  &  0.01  &      \\
  Al I & 6  &   0.03 & 0.06  &  0.13 & 0.09  &  0.06  &0.11  \\
  Si I & 11 &   0.04 & 0.14  &  0.07 & 0.13  &  0.05  &0.14  \\
  Ca I & 9  &   0.00 & 0.11  &--0.04 & 0.13  &--0.03  &0.17  \\
  Sc I & 1  &   0.06 & 0.07  &  0.03 & 0.04  &  0.01  &0.10  \\
  Sc II& 6  &   0.02 & 0.10  &  0.02 & 0.08  &  0.05  &0.08  \\
  Ti I & 16 &   0.00 & 0.08  &  0.03 & 0.09  &  0.00  &0.07  \\
  V I  & 7  &   0.03 & 0.08  &  0.10 & 0.06  &  0.13  &0.10  \\
  Cr I & 6  & --0.02 & 0.07  &  0.04 & 0.07  &  0.04  &0.13  \\
  Mn I & 2  &   0.07 & 0.12  &  0.02 & 0.09  &        &      \\
  Fe I & 26 & --0.05 & 0.04  &  0.01 & 0.06  &--0.03  &0.08  \\
  Fe II& 4  & --0.05 & 0.05  &  0.01 & 0.06  &--0.03  &0.01  \\
  Co I & 3  & --0.02 & 0.05  &  0.00 & 0.07  &  0.02  &0.08  \\
  Ni I & 17 & --0.05 & 0.09  &  0.01 & 0.09  &  0.01  &0.08  \\
  Rb I & 1  &   0.02 &       &--0.02 &       &        &      \\
  Zr I & 5  & --0.17 & 0.09  &--0.14 & 0.10  &--0.19  &0.09  \\
  Ba II& 3  & --0.04 & 0.02  &  0.01 & 0.03  &--0.05  &0.00  \\
  La II& 1  & --0.11 &       &--0.14 &       &--0.10  &      \\
             \noalign{\smallskip}
   \cline{1-8}
         \end{tabular}
      \]
   \end{table}

\subsection{Estimation of uncertainties}

The sources of uncertainties can be divided into two categories. The first 
category includes the errors which act on a single line (e.g.\ random errors 
in equivalent widths, oscillator strengths), i.e.\ uncertainties of the 
line parameters. The second category includes the errors which affect all 
the lines together, i.e.\ mainly the model errors (such as errors in the 
effective temperature, surface gravity, microturbulent velocity, etc.). 
The scatter of the deduced line abundances $\sigma$, presented in Table~4 
and 5, gives an estimate of the uncertainty coming from the random errors in 
the line parameters. The mean values of  $\sigma$=$0.08$ and 
$\sigma$=$0.13$ are for abundances derived from spectra with $R2$ and $R1$, 
accordingly. Thus the uncertainties on the derived abundances, which are the 
result of random errors, amount to approximately these values. There is a 
small systematic difference between the equivalent widths measured with the 
two cameras, however the abundance effect is small. Typically 0.03~dex 
higher abundances are obtained from the lower resolution spectra. 
  
Typical internal error estimates for the atmospheric parameters are: 
$\pm~100$~K for $T_{\rm eff}$, $\pm 0.3$~dex for log~$g$ and 
$\pm 0.3~{\rm km~s}^{-1}$ for $v_{\rm t}$. The sensitivity of the abundance 
estimates to changes in the atmospheric parameters by the assumed errors is 
illustrated  for the star F141 (Table~6). It is seen that possible 
parameter errors do not affect the abundances seriously; the element-to-iron 
ratios, which we use in our discussion, are even less sensitive. 

Since abundances of C, N and O are bound together by the molecular equilibrium 
in the stellar atmosphere, we have investigated also how an error in one of 
them effect the abundance determination of an other. 
The $\Delta{\rm [O/H]}=-0.10$ causes 
$\Delta{\rm [C/H]}=-0.04$ and $\Delta{\rm [N/H]}=0.10$, the  
$\Delta{\rm [C/H]}=-0.10$ causes $\Delta{\rm [N/H]}=0.14$ and 
$\Delta{\rm [O/H]}=-0.03$. The $\Delta {\rm [N/H]}=-0.10$ has no effect
on either the carbon nor the oxygen abundances. 
      
Other sources of observational errors, such as continuum placement or 
background subtraction problems are partly included in the equivalent width 
uncertainties discussed at the beginning of this section. 

   \begin{table}
   \setcounter{table}{5}
      \caption{Effects on derived abundances resulting from model changes 
for the star F141. The table entries show the effects on the 
logarithmic abundances relative to hydrogen, $\Delta$[A/H]. Note that the 
effects on ``relative" abundances, for example [A/Fe], are often 
considerably smaller than abundances relative to hydrogen, [A/H] } 
      \[
         \begin{tabular}{lrrc}
            \hline
            \noalign{\smallskip}
 Ion & ${ \Delta T_{\rm eff} }\atop{ -100 {\rm~K} }$ & 
            ${ \Delta \log g }\atop{ +0.3 }$ & 
            ${ \Delta v_{\rm t} }\atop{ +0.3 {\rm km~s}^{-1}}$ \\ 
            \noalign{\smallskip}
            \hline
            \noalign{\smallskip}
   C  I  &  0.02 &  0.03 &  0.00 \\
   N  I  &--0.07 &  0.01 &  0.00  \\
   O  I  &  0.01 &--0.05 &--0.01  \\  
   Na I  &--0.09 &  0.00 &--0.11  \\  
   Mg I  &--0.03 &--0.01 &--0.07  \\
   Al I  &--0.03 &  0.01 &--0.06  \\  
   Si I  &  0.05 &  0.04 &--0.05  \\    
   Ca I  &--0.11 &  0.01 &--0.12  \\
   Sc I  &--0.11 &--0.01 &--0.04  \\  
   Sc II &  0.02 &--0.05 &--0.09  \\  
   Ti I  &--0.16 &  0.00 &--0.07  \\    
   Ti II &  0.01 &  0.07 &--0.11  \\ 
   V  I  &--0.15 &  0.00 &--0.10  \\  
   Cr I  &--0.10 &  0.00 &--0.08  \\
   Mn I  &--0.07 &  0.00 &--0.13  \\         
   Fe I  &--0.05 &  0.02 &--0.11  \\
   Fe II &  0.12 &--0.06 &--0.09  \\
   Co I  &--0.05 &--0.02 &--0.03  \\
   Ni I  &--0.01 &--0.02 &--0.10  \\
   Cu I  &--0.06 &  0.02 &--0.16  \\
   Rb I  &--0.11 &  0.00 &--0.02  \\
   Y  I  &  0.01 &  0.09 &--0.12  \\
   Zr I  &--0.18 &  0.00 &--0.02  \\
   Ba II &--0.02 &--0.02 &--0.19  \\
   La II &--0.02 &--0.04 &--0.03  \\
   Ce II &--0.01 &  0.08 &--0.07  \\
   Sm II &--0.02 &  0.08 &--0.10  \\
   Eu II &  0.00 &  0.10 &--0.01  \\
                 \noalign{\smallskip}
            \hline
         \end{tabular}
      \]
   \end{table}

\section{Relative abundances}
  \begin{figure*}
    \resizebox{\hsize}{!}{\includegraphics{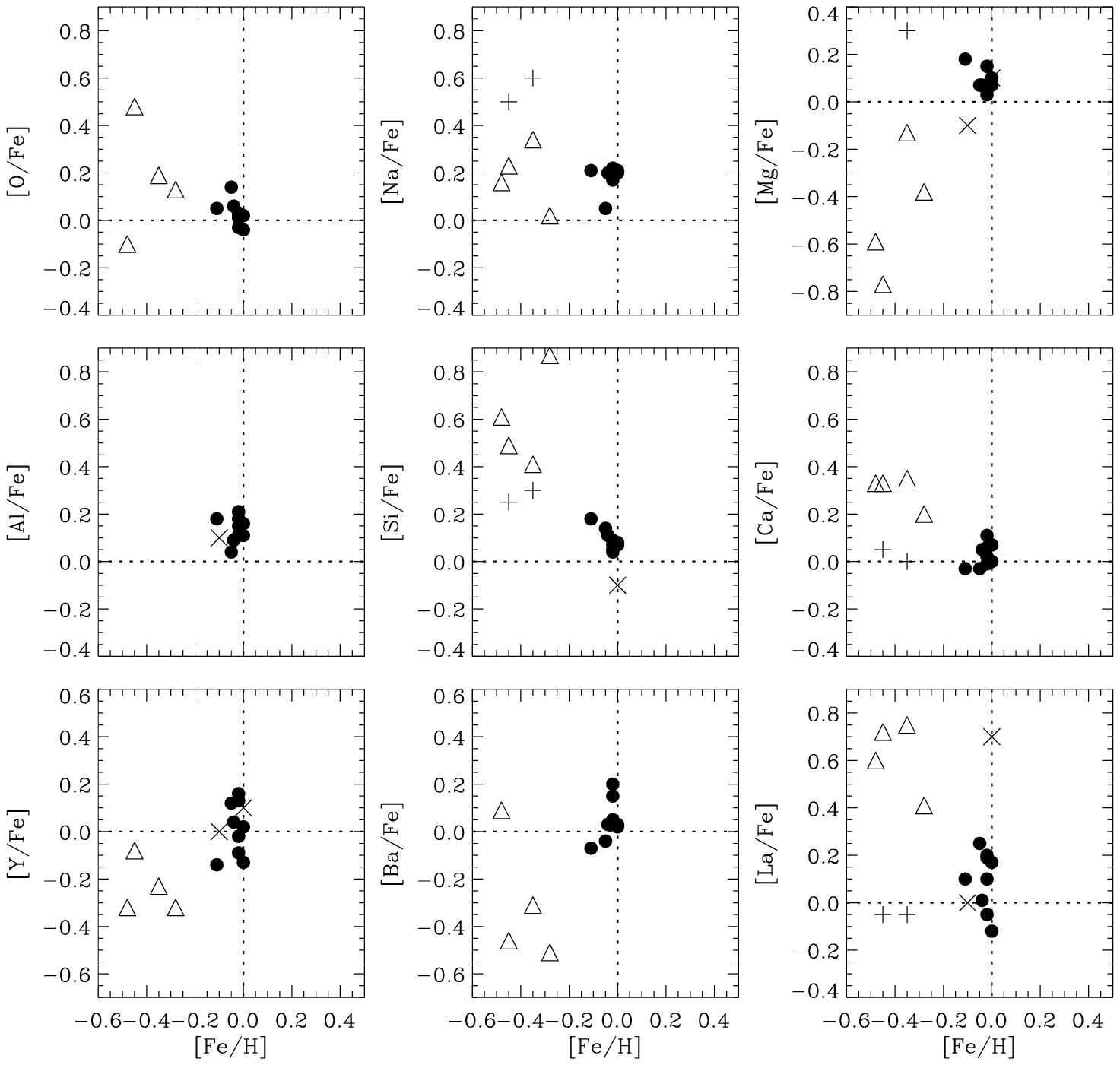}}
    \caption {Element to iron ratios as a function of iron [Fe/H].
    Results of this paper are indicated by filled circles, from
    Griffin (1975, 1979) by `plus' signs, from Cohen (1980) by
    triangles and from Foy \& Proust (1981) by crosses}
  \end{figure*}

The mean abundance of iron, [Fe/H]=$-0.03\pm0.03$, is in close 
agreement with other spectroscopic and photometric determinations: 
[Fe/H]=$-0.05$ (Canterna et al.\ 1986), [Fe/H]=$0.06$ (Nissen et al.\ 1987), 
[Fe/H]=$-0.07$ (Anthony-Twarog 1987), [Fe/H]=$0.04$ (Garcia Lopez et al.\ 
1988), [Fe/H]=$-0.08$ (Friel \& Janes 1991), [Fe/H]=$0.02$ (Friel \& 
Boesgaard 1992), [Fe/H]=$-0.09$ (Friel \& Janes 1993). 
Along with iron, the abundances of other heavy chemical elements are 
also very close to solar, as it ought to be in a cluster of almost the 
same age as the Sun (4.0~Gyr, Dinescu et al.\ 1995; 4.3~Gyr, Carraro et 
al.\ 1996; 4.0~Gyr, Boyle et al.\ 1998) and located only about 800~pc 
from it.  The small underabundance of zirconium is probably caused by 
effects of the hyperfine structure. 
 
In Fig.~2 we display 
the relative abundances of some chemical elements (for stars F84, F141 
and F151, the mean values are plotted with double weight for results 
obtained from the $R2$ spectra). We also display results 
published for M~67 by other authors. A detailed 
spectroscopic analysis for IV-202 was done by Griffin (1975) and for T626 by 
Griffin (1979), for four stars (F105, F170, F224, and F231) by Cohen (1980) 
and for F164 and F170 by Foy \& Proust (1981). The large 
scatter of the abundance ratios in these early analyses, when present, is 
most probably due to observational errors caused by photographic data used. 
    
In our study, among mixing-sensitive chemical elements alterations of 
carbon, nitrogen and sodium are noticeable. 
In Sects. 5.1 and 5.2 we describe them in more detail. 
 
\subsection{Carbon and nitrogen}

There are several possibilities to investigate carbon abundances in stars: 
a low excitation [C~I] 
line at 8727~{\AA}, a number of high excitation C~I lines and molecular lines 
of CH and ${\rm C}_2$. Due to its low excitation potential, the [C~I] line 
should not be sensitive to non-LTE effects and to uncertainties in the 
adopted model atmosphere parameters, contrary to what may be expected for the 
frequently used high excitation C I and CH lines. As it can be seen from 
results by Clegg et al. (1981), carbon abundances obtained using 
the high excitation C~I lines and CH molecular lines are systematically 
lower than abundances obtained from the 8727~{\AA} [C~I] line. The agreement 
is present only for results obtained from ${\rm C}_2$ molecular lines. A 
comprehensive discussion on this subject is presented by Gustafsson et al. 
(1999) and Samland (1998). 

Unfortunately, the [C~I] 8727~{\AA} line is not suitable for the analysis 
in spectra of red giants, especially when a probability of increased 
strengths of CN molecular lines is present. Lambert and Ries (1977) examined 
CNO abundances in K-type giants and found that the [C~I] line was not 
trustworthy as a C abundance indicator, devoting an Appendix to the problem. 
Consequently, we decided to use the ${\rm C}_2$ Swan (0,1) band head 
at 5635.5~{\AA} for the analysis. This feature is strong enough in our 
spectra and is quite sensitive to changes of the carbon abundance (see 
Fig.~3 for illustration).     
 
  \begin{figure}
    \resizebox{\hsize}{!}{\includegraphics{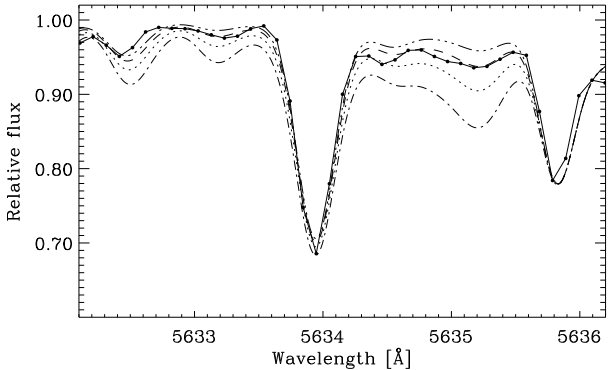}}
    \caption{
    Synthetic (dashed and dotted curves) and observed (solid curve and dots)
    spectra for the 1--0 C$_2$ region near $\lambda 5635$ of F151. The 
    syntheses were generated with [C/H]=0, --0.1, --0.2, and --0.3}
  \end{figure}

The previous high-resolution determination of [C/H] in M~67 giants 
(F105, F108 and F170) was carried out by Brown (1985), using C$_2$ lines 
at around 5110~{\AA} and 4730~{\AA}, and gave the mean value [C/H]=--0.26. 
Our result for these stars is slightly higher [C/H]=--0.19$\pm 0.04$. The mean 
value for the clump stars is --0.21$\pm 0.05$~dex. 
     
The available study of carbon abundances in dwarf stars of M~67 is that by 
Friel \& Boesgaard (1992). Six high excitation C~I lines in the spectral 
region from 7100~{\AA} to 7120~{\AA} were analysed in three F dwarfs and the 
mean [C/H]=$-0.09\pm0.03$ was obtained. This value, probably, would be by 
about 0.1~dex higher if the low excitation [C~I] line would be used instead. 
Then the [C/H] value in dwarfs of M~67 would be equal to the solar value.  
Thus, compared with the Sun and with dwarf stars of M~67, the carbon abundance 
might be depleted by about 0.2~dex in the stars we analysed. 

Another evaluation of the carbon abundance can be done by the 
comparison with carbon abundances  determined for dwarf stars 
in the galactic disk. Gustafsson et al.\ (1999), using the forbidden 
[C~I] line, have performed an abundance analysis of carbon in a sample of 80 
late F and early G type dwarfs. As is seen from Fig.~4, 
the ratios of [C/Fe] and [C/O] in our stars lie by about 0.2~dex below 
the trends obtained for dwarf stars in the galactic disk.      

  \begin{figure}
    \resizebox{\hsize}{!}{\includegraphics{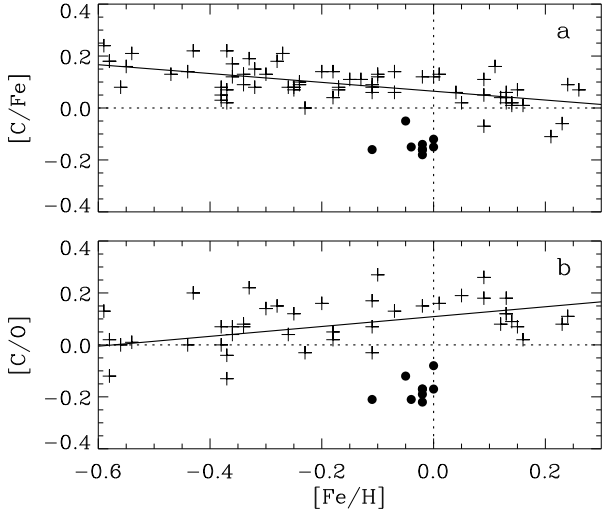}}
    \caption{[C/Fe] ({\bf a}) and [C/O] ({\bf b}) as a function of [Fe/H]. 
    Results of this paper are indicated by filled circles, results obtained for
    dwarf stars of the galactic disk (Gustafsson et al.\ 1999) by `plus' signs
    and the full line}
  \end{figure}

The wavelength 
interval 7980--8130~{\AA}, with 65 CN lines selected, was analysed 
in order to determine the nitrogen abundances. The mean nitrogen abundance, 
as determined from the giants, is [N/H]=$0.18\pm0.02$ and from the clump 
stars [N/H]=$0.24\pm0.08$. Neither the carbon depletion, nor the nitrogen 
enrichment are as large as was reported by Brown (1985).  Consequently, 
the C/N ratios obtained in our work do not 
request large extra-mixing processes in order to be explained.    
The mean C/N ratios are lowered to the value of 1.72 in the giants 
and to the values of 1.44 in the clump stars. 

In our work, $^{12}{\rm C}/^{13}{\rm C}$ ratios were determined for all
programme stars from the (2,0)
$^{13}{\rm C}^{14}{\rm N}$ line at 8004.728~{\AA} with a laboratory wavelength
adopted from Wyller (1966). 
Fig.~5 illustrates the enhancement of $^{13}{\rm CN}$ line at 8004~{\AA} in 
spectra of F164 and F224. The star F224 has the lowest value 
$^{12}{\rm C}/^{13}{\rm C}$=8, among stars investigated in our work. 
For F84, F141 and F151, the $^{13}{\rm CN}$ line at
8381.06~{\AA} were observed at higher resolution, but due to the weakness of
the line we can for F84 and F141 only confirm that the
$^{12}{\rm C}/^{13}{\rm C}$ ratios are consistent with the values derived from
the 8004.7~{\AA} feature, while for F151 we are not able to derive any
useful information of the isotope ratio from this line.
 
  \begin{figure}
    \resizebox{\hsize}{!}{\includegraphics{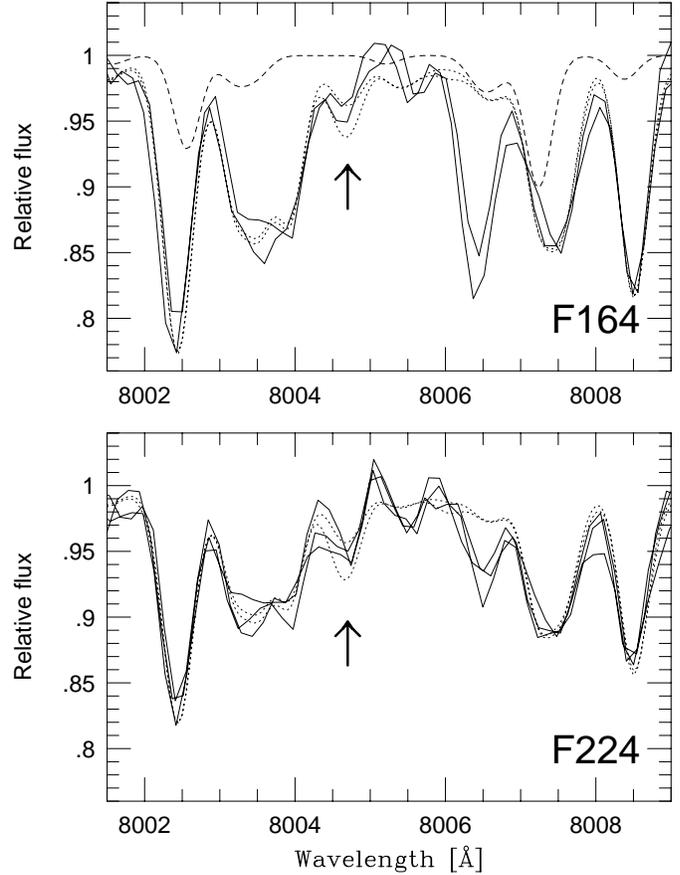}}
    \caption{
    A small portion of the 8000~\AA\ wavelength interval showing the
    8004.7 \AA\ $^{13}$CN feature in F164 and F224.
    In the upper panel, the solid lines show the 2 observed spectra for F164,
    the dotted lines show two synthetic spectra with $^{12}$C/$^{13}$C
    ratios of 10 and 20, the dashed line shows a synthetic spectrum without 
    any CN lines.
    In the lower panel, the solid lines show the 3 observed spectra for F224,
    and the dotted lines show two synthetic spectra with
    $^{12}$C/$^{13}$C ratios of 5 and 10}
    \label{CN}
  \end{figure}

  \begin{figure}
    \resizebox{\hsize}{!}{\includegraphics{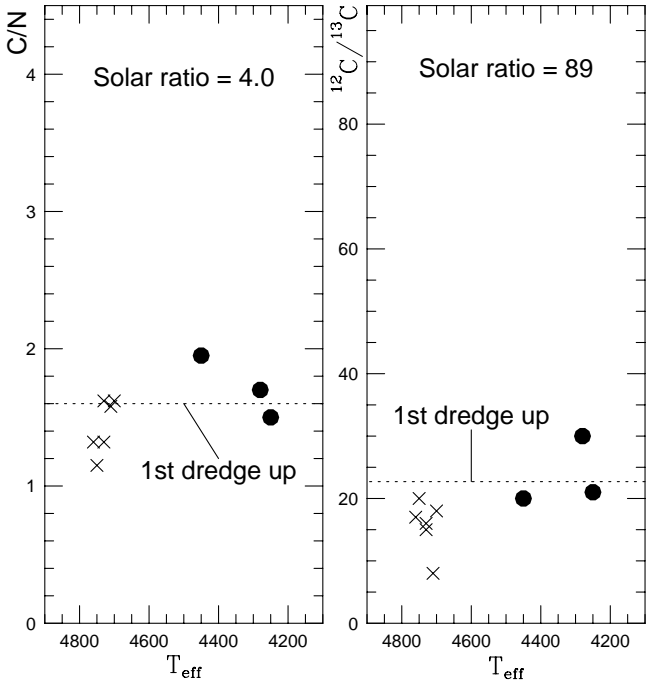}}
    \caption{C/N and $^{12}$C/$^{13}$C abundance ratios for giants
    ({\it filled circles}) and clump stars ({\it crosses}).
    The dotted lines show predictions from Charbonnel (1994).
    We suggest that the diagram shows that extra mixing
    takes place after the He-core flash
    }
  \end{figure}

Ratios of $^{12}{\rm C}/^{13}{\rm C}$ were investigated for stars in 
M~67 by Gilroy (1989) and Gilroy \& Brown (1991). There are six stars in 
common with Gilroy (1989) and one with Gilroy \& Brown (1991). Except 
for two stars (F84 and F170), our $^{12}{\rm C}/^{13}{\rm C}$ ratios agree 
within errors of uncertainties. The mean difference between our values and 
these by Gilroy \& Brown is equal to $3\pm4$. 
In their study, Gilroy \& Brown rule out mixing during the He-core flash 
because the two stars, F108 and F170, had $^{12}{\rm C}/^{13}{\rm C}$ ratios 
similar to the clump stars.  In our study however, a small difference can 
be suspected. We find the mean $^{12}{\rm C}/^{13}{\rm C}$ ratios lowered 
to the value of $24\pm4$ in the giants and to the value of $16\pm4$ in the 
clump stars.  

The standard theoretical evolution of the carbon isotopic ratio and carbon 
to nitrogen ratio along the giant branch was homogeneously mapped by 
Charbonnel (1994) for stellar masses between 1 and $7~M_{\odot}$, and for 
different values of metallicity. For a 1.25~$M_{\odot}$ star (approximately
the turn-off mass of M~67), with initially solar composition the 
predicted $^{12}{\rm C}/^{13}{\rm C}$ and $^{12}{\rm C}/^{14}{\rm N}$ ratios 
at the end of the first dredge-up phase are about 22.7 and 1.6, respectively 
(Charbonnel 1994, Figs. 2 and 4 (it is not explaned in the paper why values 
in Table 2 and Figs. 7 and 8 are different from those presented in Figs. 2 and 
4, so we decided to use homogeneous ones)).
The predicted values are in good agreement with our results for the giant
stars. The clump stars, however, may well show an additional decrease caused
by an extra mixing process (see Fig. 6).

\subsection{Sodium}
The stars in our sample, as determined from the Na~I lines $\lambda$~5682.64, 
6154.23 and 6160.75~\AA\, show a slight overabundance of sodium (see Fig.~7).     

Overabundances of sodium in red giants have long been considered as being of 
a primordial origin (see, e.g., Cottrell \& Da Costa 1981). However, the 
star-to-star variations of Na, the existence of Na versus N correlations, 
and Na versus O anticorrelations in globular cluster red giants have revealed 
a possibility of sodium to be produced in red giant stars (Cohen 1978, 
Peterson 1980, Drake et al.\ 1992, Kraft et al.\ 1995, 1997 and references 
therein). Norris \& Da Costa (1995) have concluded that Na variations exist in 
all clusters, while Al variations are greater in the more metal-poor clusters. 
Theoretical explanations for the production of Na and Al have been proposed 
by Sweigart \& Mengel (1979), Denissenkov \& Denissenkova (1990), 
Langer \& Hoffman (1995), Cavallo et al.\ (1996) and other studies; however, 
the origin and extent of the phenomenon is not well understood.

For the stars in our sample, the overabundance of sodium is not followed by 
a noticeable overabundance of aluminium and underabundance of oxygen. 
This confirms the conclusion by Norris \& Da Costa (1995) that Al variations 
are not that great as of Na in metal-rich stars.  
The overabundance of Na could appear due to the deep mixing from layers of 
the NeNa cycle, which lie higher than ON-processed regions in red giants 
(cf.\ Cavallo et al.\ 1996). Shetrone (1996), from the analysis of red giants 
in the globular cluster M~71, also concluded that either Al and Na are created 
in different nucleosynthesis processes, or the NeNa cycle can occur without 
the ON or MgAl cycles. 

  \begin{figure*}
    \resizebox{\hsize}{!}{\includegraphics{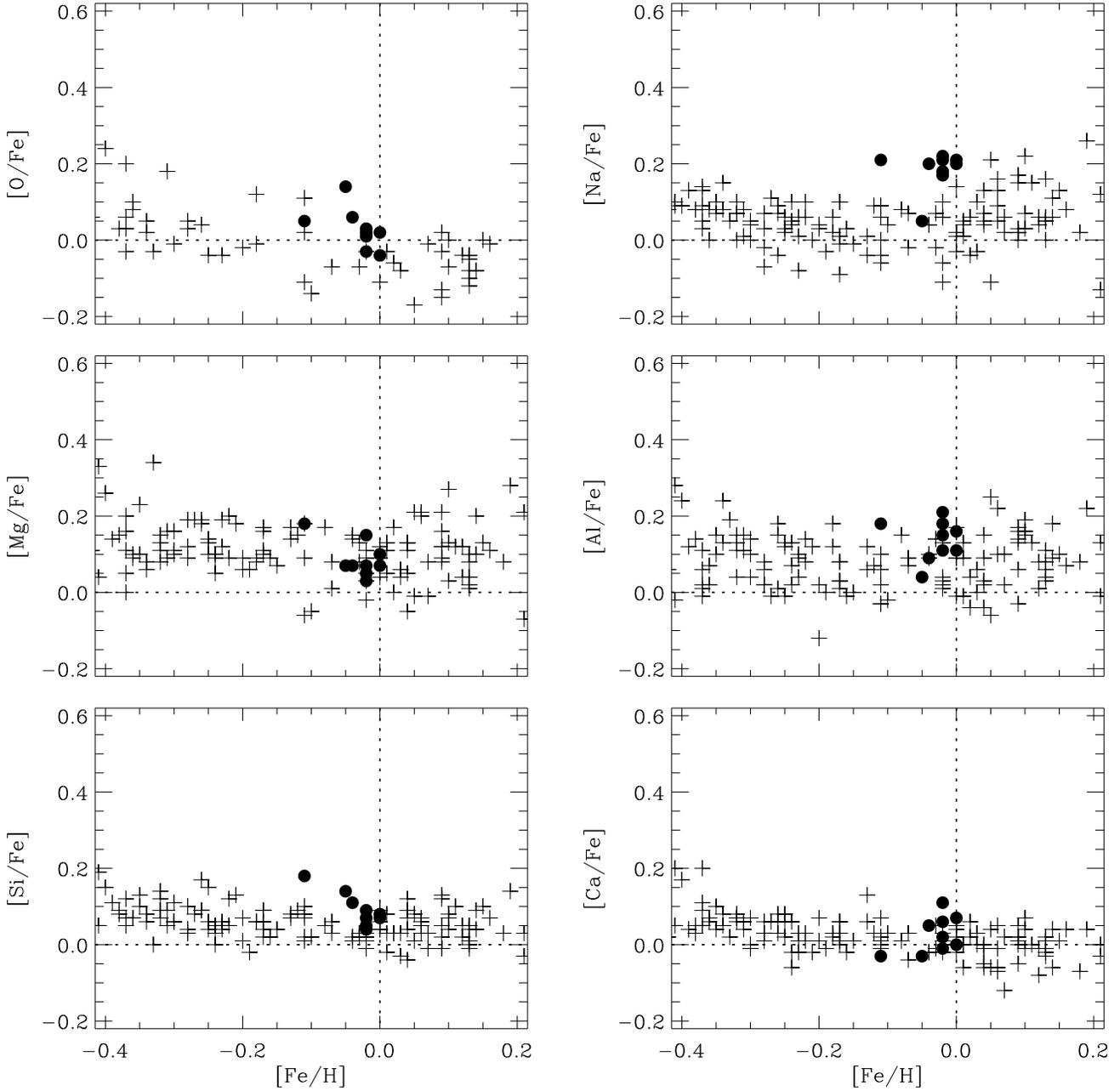}}
    \caption{Element to iron ratios as a function of iron [Fe/H]. Results of this 
    paper are indicated by {\it filled circles}, 
    results for the Galactic disk stars investigated by Edvardsson et al. (1993) 
    by {\it crosses}}
  \end{figure*}

\subsection{Final remarks}

The change in the surface composition of a star ascending the giant branch 
is predicted by theoretical calculations. When a star evolves up the giant 
branch its convective envelope deepens and CN-cycle products are mixed to 
the surface of the evolving star, causing the surface 
$^{12}{\rm C}/^{13}{\rm C}$ and $^{12}{\rm C}/^{14}{\rm N}$ ratios to drop. 
These ratios decrease with increasing stellar mass and decreasing metallicity. 
Extra-mixing processes may become efficient on the red giant branch when 
stars reach the so-called luminosity function bump and modify the surface 
abundances (see Charbonnel et al.\ 1998 for more discussion). 
In case of M~67, this may happen starting from log$L/L_{\odot}$=1.64 (Charbonnel 
1994), however the first and only evidence on the evolutionary state at 
whish this non-standard mixing actually becomes effective has to come from 
observations. The giants F108 and F170 (log$L/L_{\odot}\approx2.3$), 
observed in our work, do not show obvious effects of the extra mixing. Other 
bright M~67 giants such as T626 and IV-202 are not investigated since have 
quite low membership probabilities (19\% and 51\%, respectively, as quoted by 
Sanders 1977) and are rather cool (continuum placement in their spectra would 
be at best difficult). In our work, the extra-mixing processes seem to show
up in the clump stars observed (Fig. 6), however here the He-core flash may
be responsible.
   
The role that the He-core flash may play in producing surface abundance 
changes still has to be investigated. The theoretical 
calculations indicate that the nature of nucleosynthesis and mixing depend 
upon the degree of degeneracy in the He-core and, hence, intensity of the 
explosion: intermediate flashes produce the most mixing 
(Despain 1982, Deupree 1986, Deupree \& Wallace 1987, Wallace 1987 and 
references therein). Due to the numerical difficulties in treating such a 
violent event, the He core-flash remains an event of interest to theorists. 
The observational data for giants and clump stars of M~67 in our work show 
a slight increase of abundance changes in more evolved clump stars. New 
precise observational data are necessary in understanding effects of the 
He-core flash and other open questions of chemical evolution of stars. 

\begin{acknowledgements}
We are much indebted to the staff of the NOT 
for they willing and capable help during the observing runs. Heidi Kor\-honen 
(NOT) and Eduardas Puzeras (ITPA) are thanked for their help in spectral 
reductions.
Bertrand Plez (University of Montpellier II) and Guillermo Gonzalez  
(Washington State University) were particularly generous in providing us with 
atomic data for CN and C$_2$ molecules, respectively.
The helpful comments to the manuscript by Rudolf Duemmler (Oulu University) 
and by the referee Jeffery Brown (Washington State University) are 
appreciated. G.T. is grateful to Per Lilje (Oslo University) for supporting 
the observing trip to the NOT. Information from the Vienna Atomic Line Data 
Base was very useful in compiling atomic line data. 
B.E. is supported by the Swedish Natural Sciences Research Council (NFR).
\end{acknowledgements}

\end{document}